\documentstyle[12pt]{article}

\begin{document}
\baselineskip 0.7cm

\newcommand{\gsim}{ \mathop{}_{\textstyle \sim}^{\textstyle >} }
\newcommand{\lsim}{ \mathop{}_{\textstyle \sim}^{\textstyle <} }
\newcommand{\EV}{ {\rm eV} }
\newcommand{\KEV}{ {\rm keV} }
\newcommand{\MEV}{ {\rm MeV} }
\newcommand{\GEV}{ {\rm GeV} }
\newcommand{\TEV}{ {\rm TeV} }
\renewcommand{\thefootnote}{\fnsymbol{footnote}}
\setcounter{footnote}{1}

\begin{titlepage}
%\today
\begin{flushright}
UT-823
\\
July 1998
\end{flushright}

\vskip 0.35cm
\begin{center}
{\large \bf 
Bi-Maximal Neutrino Mixing in SO(10)$_{\rm GUT}$
}
\vskip 1.2cm
Yasunori Nomura and T.~Yanagida
\vskip 0.4cm

{\it Department of Physics and RESCEU, University of Tokyo,\\
     Bunkyo-ku, Hongou, Tokyo 113, Japan}

\vskip 1.5cm

\abstract{We find a grand unified SO(10) model which accommodates 
the bi-maximal neutrino mixing for vacuum-oscillation solutions 
to the atmospheric and the solar neutrino problems.
This model maintains the original SO(10) mass relation between 
neutrino and up-type quark masses, 
$m_{\nu_2} / m_{\nu_3} \sim (m_{c}/m_{t})^2$.}
\end{center}
\end{titlepage}

\renewcommand{\thefootnote}{\arabic{footnote}}
\setcounter{footnote}{0}

%
%
%       *** Main Part ***
%
%

The recent data on the atmospheric neutrino from the SuperKamiokande 
(SuperK) Collaboration \cite{Super_K_atm} have presented convincing
evidence for neutrino oscillation with mass-squared difference 
$\delta m_{\rm atm}^2 \simeq 5 \times 10^{-3}~\EV^2$.
It is now understood that the long-standing puzzle of the atmospheric
muon neutrino ($\nu_{\mu}$) deficit in underground detectors 
\cite{nu_mu_deficit} is indeed due to the neutrino oscillation.
As for the solar neutrino problem, there are still two allowed solutions 
: one is matter enhanced neutrino oscillation 
({\it i.e.} the MSW solution \cite{MSW})\footnote{
The MSW solution has two distinct regions : the small and the large
angle ones \cite{Hata_Langacker}.} 
and the other is long-distance vacuum neutrino
oscillation called as ``just-so'' oscillation \cite{BPW, GK}.

It is known \cite{MSW_atm_Seesaw} that the small angle MSW solution and
the maximal mixing between the atmospheric $\nu_{\mu}$ and $\nu_{\tau}$
are quite naturally explained in a large class of see-saw models
\cite{Seesaw_models}.
However, the electron energy spectrum recently reported by the SuperK
Collaboration \cite{Super_K_sun} seems to favor the ``just-so'' vacuum
oscillation with $\delta m_{\rm sun}^2 \simeq 10^{-10}~\EV^2$ and the
maximal mixing.
If this vacuum oscillation of the solar neutrino is confirmed in future
solar-neutrino experiments \cite{Super_K_sun, SNO}, 
we are led to a quite surprising situation that two
independent mixing angles in the lepton sector are very large in
contrast with the quark sector in which all observed mixing angles among 
different families are small.
This may point to that a rule governs the lepton mass matrices is
significantly different from the one for the quark sector, which seems
to be a contradiction to the idea of complete unification of quarks and
leptons.\footnote{
The so-called democratic mass matrices for quarks and leptons can
generate large mixing in the neutrino sector \cite{democratic}.}

On the other hand, as noted recently by Barger et.al.~\cite{BPWW} the
required neutrino mass ratio $m_{\nu_2} / m_{\nu_3} \simeq 10^{-4}$ 
(provided $m_{\nu_1} \ll m_{\nu_2} \ll m_{\nu_3}$) is approximately 
equal to $(m_{c}/m_{t})^2$ as predicted by a see-saw model in the
SO(10) grand unified theory (GUT) \cite{Seesaw}.

The purpose of this paper is to construct a simple SO(10)$_{\rm GUT}$
model which naturally accomodates the bi-maximal neutrino mixing
for the atmospheric and the solar neutrino vacuum oscillations,
keeping the interesting SO(10)$_{\rm GUT}$ mass relation 
$m_{\nu_2} / m_{\nu_3} \sim (m_{c}/m_{t})^2$.\footnote{
Recent analyses on phenomenological consequences of the bi-maximal
neutrino mixing are given in Ref.~\cite{BPWW, BGG}.}
We assume supersymmetry throughout this paper.

Let us first discuss the minimal SO(10)$_{\rm GUT}$ model which contains 
three families of quarks and leptons $\psi_i({\bf 16})$ $(i=1,\cdots,3)$ 
belonging to ${\bf 16}$ of the SO(10)$_{\rm GUT}$ and one Higgs field 
$H({\bf 10})$.
We will consider Higgs multiplets responsible for the breaking of
SO(10)$_{\rm GUT}$ down to the standard-model gauge group later.
This minimal model is known to yield a mass degeneracy of up-type and
down-type quarks and vanishing CKM mixing \cite{Langacker}.

The simplest extension of the minimal model avoiding this unwanted mass
degenercy is to introduce another Higgs field $H'({\bf 10})$.
This two-Higgs field ${\bf 10}$ model, in fact, gives less stringent
relations among quark and lepton mass matrices as
\begin{eqnarray}
  M_{\nu D} = M_u \qquad , \qquad M_l = M_d,
\end{eqnarray}
where $M_{\nu D}$ is 3$\times$3 Dirac mass matrix for neutrinos.
It is well known \cite{YY} that we need a large hierarchy in the Majorana
mass matrix for right-handed neutrinos $N_i$ $(i=1,\cdots,3)$ to obtain
large neutrino mixing.
However, if one assumes that the large hierarchy in the Majorana mass
matrix for $N_i$ one loses the original SO(10)$_{\rm GUT}$ relation, 
$m_{\nu_2} / m_{\nu_3} \sim (m_{c}/m_{t})^2$, discussed in the
introduction.

We, therefore, consider a different extension of the minimal
SO(10)$_{\rm GUT}$ model in this paper.
Instead of adding an extra Higgs field $H'({\bf 10})$, we introduce one
extra matter multiplet $\psi({\bf 10})$ belonging to ${\bf 10}$ of the 
SO(10)$_{\rm GUT}$ \cite{HMY}.
Thus, the matter multiplets in our model are three families of 
$\psi_i({\bf 16})$ and one $\psi({\bf 10})$.\footnote{
We introduce one extra matter multiplet $\psi({\bf 10})$ in this paper.
We may, however, introduce three families of $\psi_i({\bf 10})$ 
$(i=1,\cdots,3)$.
In this case, the Froggatt-Nielsen mechanism \cite{Froggatt_Nielsen} may 
be used to account for observed quark and lepton mass matrices, assuming 
different charges for $\psi_i({\bf 16})$ and $\psi_i({\bf 10})$ in each
families.}

We now assume that the SO(10)$_{\rm GUT}$ is broken down to 
SU(5) by condensation of Higgs fields 
$\langle \chi({\bf 16}) \rangle = \langle \bar{\chi}({\bf 16}^*) \rangle = V$  
with $V$ being $\sim 10^{16}~\GEV$.\footnote{
We need other Higgs multiplets such as ${\bf 45}$ to complete the
breaking of the SO(10)$_{\rm GUT}$ down to the standard-model gauge
group.
We do not consider them in this paper, since they are irrelevant to our
present analysis.}
This GUT breaking also induces a mass term for the matter multiplets
through the following superpotential,
\begin{eqnarray}
  W = \sum_{i=1}^{3} f_i\, \psi_i({\bf 16})\, \psi({\bf 10}) 
      \langle \chi({\bf 16}) \rangle.
\label{mass_matter}
\end{eqnarray}
Namely, a linear combination ${{\bf 5}^{*}}'_{\psi} \equiv
\sum_{i=1}^{3} f_i\, {\bf 5}^*_i$ in $\psi_i({\bf 16})$ receives a GUT
scale mass together with ${\bf 5}_{\psi}$ in $\psi({\bf 10})$.
We choose $f_2 = f_3 = 0$, here.
The reason for this will be clearly understood later on.

After the spontaneous breakdown of SO(10)$_{\rm GUT}$ to SU(5), massless 
matter multiplets are given by
\begin{eqnarray}
\begin{array}{ccccccc}
  {\bf 10}_3 &+&   {\bf 5}^*_3   &+& N_3 &   =   & \psi_3({\bf 16}), \\
  {\bf 10}_2 &+&   {\bf 5}^*_2   &+& N_2 &   =   & \psi_2({\bf 16}), \\
  {\bf 10}_1 & &                 &+& N_1 &\subset& \psi_1({\bf 16}), \\
             & & {\bf 5}^*_{\psi} &&     &\subset& \psi({\bf 10}).
\end{array}
\label{content}
\end{eqnarray}
It should be clear that ${\bf 10}_i$ $(i=1,\cdots,3)$ in 
Eq.~(\ref{content}) are not ${\bf 10}$ of the SO(10)$_{\rm GUT}$, 
but ${\bf 10}$ of the SU(5).

We take a basis where the original Yukawa coupling matrix of the Higgs
field $H({\bf 10})$ to the matter $\psi_i({\bf 16})$ is
diagonal;\footnote{
Precisely speaking, we assume $f_2 = f_3 = 0$ in Eq.~(\ref{mass_matter}) 
in this basis.
We discard small deviations from our assumption in the present analysis.}
\begin{eqnarray}
  W = h_i\, \psi_i({\bf 16})\, \psi_i({\bf 16})\, H({\bf 10}).
\label{diagonal_Yukawa}
\end{eqnarray}
This leads to a diagonal mass matrx for the up-type quarks as
\begin{eqnarray}
  M_u = \pmatrix{
          m_u &  0  &  0  \cr
           0  & m_c &  0  \cr
           0  &  0  & m_t \cr},
\end{eqnarray}
where $M_u$ is defined as
\begin{eqnarray}
  (M_u)_{ii} = h_i\, \langle {\bf 5}_H \rangle.
\end{eqnarray}
Here, ${\bf 5}_H$ is a SU(5)-${\bf 5}$ component of 
$H({\bf 10})$.
 
The down-type quark mass matrix is, however, incomplete, since the 
SU(5)-${\bf 5}^*$ of $\psi({\bf 10})$ ({\it i.e.} ${\bf 5}^*_{\psi}$),
does not have any Yukawa coupling to $H({\bf 10})$.
To solve this problem we introduce a pair of Higgs fields $H({\bf 16})$
and $\bar{H}({\bf 16}^*)$ and consider a superpotential
\begin{eqnarray}
  W = k\, H({\bf 10})\, \bar{H}({\bf 16}^*)\, \bar{\chi}({\bf 16}^*) 
      + M\, H({\bf 16})\, \bar{H}({\bf 16}^*).
\end{eqnarray}
U(1) $R$-symmetry may be useful to have this form of superpotential.
The U(1)$_R$ charges are given in Table~\ref{R_charge}.
\begin{table}
\begin{center}
\begin{tabular}{|c|ccccccc|}  \hline 
  & $H({\bf 10})$ & $H({\bf 16})$ & $\bar{H}({\bf 16}^*)$ & 
    $\chi({\bf 16})$ & $\bar{\chi}({\bf 16}^*)$ & $\psi_i({\bf 16})$ &
    $\psi({\bf 10})$ \\ \hline
  $R$ & 0 & 0 & 2 & 0 & 0 & 1 & 1 \\ \hline
\end{tabular}
\end{center}
\caption{U(1)$_R$ charges.}
\label{R_charge}
\end{table}
The GUT condensation $\langle \bar{\chi}({\bf 16}^*) \rangle \neq 0$
induces a mass mixing between ${\bf 5}^*$'s of $H({\bf 10})$ and 
$H({\bf 16})$ ({\it i.e.} ${\bf 5}^*_{H({\bf 10})}$ and 
${\bf 5}^*_{H({\bf 16})}$).
Then, a linear combination
\begin{eqnarray}
  \tilde{\bf 5}^*_H &=& \cos\theta\, {\bf 5}^*_{H({\bf 10})}
                      + \sin\theta\, {\bf 5}^*_{H({\bf 16})},\\
  \tan\theta &=& - \frac{k\, \langle \bar{\chi}({\bf 16}^*) \rangle}{M},
\end{eqnarray}
remains as a massless Higgs field $H({\bf 5}^*)$ in the standard 
SU(5)$_{\rm GUT}$ and contributes to the quark and lepton mass matrix.
Then, $\tilde{\bf 5}^*_H$ can couple to ${\bf 5}^*_{\psi}$ as
\begin{eqnarray}
  W_{\rm eff} = \sin\theta \sum_{i=1}^{3} g_i\, {\bf 10}_i\, 
                {\bf 5}^*_{\psi}\, \tilde{\bf 5}^*_H,
\end{eqnarray}
where the coupling constants $g_i$ are defined as
\begin{eqnarray}
  W = \sum_{i=1}^{3} g_i\, \psi_i({\bf 16})\, \psi({\bf 10})\, H({\bf 16}).
\end{eqnarray}

Now, the Yukawa coupling of $\tilde{\bf 5}^*_H$ is given by
\begin{eqnarray}
  W_{\rm eff} = \cos\theta
      \pmatrix{ {\bf 10}_1, & {\bf 10}_2, & {\bf 10}_3 \cr}
      \pmatrix{   0   & g_1 \tan\theta &  0    \cr
                h_{2} & g_2 \tan\theta &  0    \cr
                  0   & g_3 \tan\theta & h_{3} \cr}
      \pmatrix{ {\bf 5}^*_2      \cr
                {\bf 5}^*_{\psi} \cr
                {\bf 5}^*_3      \cr}
      \tilde{\bf 5}^*_H,
\end{eqnarray}
which yields the down-type quark and the charged lepton mass matrix,
\begin{eqnarray}
  M_{d/l} = m_t \pmatrix{
                  0    & x & 0  \cr
               m_c/m_t & y & 0  \cr
                  0    & z & 1  \cr}\times
            \frac{\cos\theta}{\tan\beta}.
\end{eqnarray}
Here, $\tan\beta \equiv \langle {\bf 5}_H \rangle / 
\langle \tilde{\bf 5}^*_H \rangle$ and
\begin{eqnarray}
  x = \frac{g_1}{h_{3}}\tan\theta, \qquad 
  y = \frac{g_2}{h_{3}}\tan\theta, \qquad
  z = \frac{g_3}{h_{3}}\tan\theta.
\end{eqnarray}
We see that a choice of $x \sim m_c/m_t$, $y \sim \sqrt{m_c/m_t}$ and 
$z \sim 1$ produces a nice fit of the observed quark and
lepton mass ratios and the CKM matrix.\footnote{
To explain quark and lepton masses more precisely one must introduce
SU(5) breaking effects, otherwise we have wrong SU(5)$_{\rm GUT}$ 
relations, $m_{\mu} = m_s$ and $m_e = m_d$.
A detailed analysis including these effects will be given in
Ref.~\cite{Future}.}
Thus, we take $x \simeq m_c/m_t$, $y \simeq \sqrt{m_c/m_t}$ and 
$z \simeq 1$.
The $\tan\beta$ may be very large unless $\cos\theta$ is very small
($\tan\beta \simeq \sqrt{2} (m_t/m_b) \cos\theta$).
It is now clear that in contrast with the CKM mixing we have a large
mixing closed to the maximal between ${\bf 5}^*_3$ and ${\bf 5}^*_{\psi}$ 
which corresponds to a mixing between charged leptons of the third and 
second families.

Let us turn to the Dirac mass term for neutrinos which is given by the
superpotential (\ref{diagonal_Yukawa}).
This mass matrix is also incomplete, since ${\bf 5}^*_{\psi}$ never
couples to $N_i$'s in $\psi_i({\bf 16})$.
However, the following nonrenormalizable interaction gives a desired
coupling:
\begin{eqnarray}
  W = \sum_{i=1}^{3} k_i\, \psi_i({\bf 16})\, \psi({\bf 10})\, H({\bf 10}) 
      \frac{\bar{\chi}({\bf 16}^*)}{M_G},
\label{nonren_nuD}
\end{eqnarray}
where $M_G$ is the gravitational scale $M_G \simeq 2 \times
10^{18}~\GEV$.
Together with the original coupling in Eq.~(\ref{diagonal_Yukawa}), 
the nonrenormalizable interaction (\ref{nonren_nuD}) yields
\begin{eqnarray}
  M_{\nu D} = m_t \pmatrix{
                   0    & \delta_1 &  0  \cr
                m_c/m_t & \delta_2 &  0  \cr
                   0    & \delta_3 &  1  \cr}.
\end{eqnarray}
Here, $M_{\nu D}$ is defined as
\begin{eqnarray}
  W_{\rm eff} = \pmatrix{ N_1, & N_2, & N_3 \cr}
                M_{\nu D}
                \pmatrix{ {\bf 5}^*_2      \cr
                          {\bf 5}^*_{\psi} \cr
                          {\bf 5}^*_3      \cr},
\end{eqnarray}
and $\delta_i = (k_i/h_{3})(\langle \chi({\bf 16}) \rangle/M_G)$.
Notice that $\delta_i \simeq O(10^{-2})$ as long as $k_i/h_3 \sim O(1)$.

It is extremely interesting that when $\delta_2 \sim m_c/m_t$ we have a
large mixing between ${\bf 5}^*_{\psi}$ and ${\bf 5}^*_2$ which produces 
a large mixing between left-handed neutrinos of the second and first
families.
This observation is crucial for our purpose, since this tells us that
when we maintain the SO(10)$_{\rm GUT}$ relation, $m_{\nu_2}/m_{\nu_3}
\sim (m_c/m_t)^2$ ({\it i.e.} $\delta_2 \simeq m_c/m_t$), we necessarily
obtain a large mixing closed to the maximal between $\nu_e$ and
$\nu_{\mu}$ (provided that the Majorana mass matrix for $N_i$ does not
have hierarchy).
We take, for simplicity, $\delta_1 \simeq \delta_3 \simeq 0$ and 
$\delta_2 \simeq m_c/m_t$.\footnote{
If $\delta_2 \simeq 1/5$, the small angle MSW solution can be
accommodated instead of the ``just-so'' solution.
In this case some SO(10)-singlet fields are required at the GUT scale.}

The Majorana masses for the right-handed neutrino $N_i$ are given by the 
following nonrenormalizable superpotential,
\begin{eqnarray}
  W = j_{ij} \frac{1}{M_G} \psi_i({\bf 16})\, \psi_j({\bf 16})\, 
      \bar{\chi}({\bf 16}^*)\, \bar{\chi}({\bf 16}^*).
\end{eqnarray}
After the SO(10)$_{\rm GUT}$ breaking 
we obtain the Majorana mass matrix,
\begin{eqnarray}
  (M_N)_{ij} = \frac{\langle \bar{\chi}({\bf 16}^*) \rangle^2}{M_G} (j_{ij}).
\end{eqnarray}
Simply assuming $\langle \bar{\chi}({\bf 16}^*) \rangle = V \sim
10^{16}~\GEV$ and $j_{ij} \sim \delta_{ij}$ we get
\begin{eqnarray}
  M_N \sim (10^{14}~\GEV) \times \pmatrix{
                 1 & 0 & 0  \cr
                 0 & 1 & 0  \cr
                 0 & 0 & 1  \cr}.
\end{eqnarray}
From the see-saw mechanism the light neutrino masses are given by
\begin{eqnarray}
  M_{\nu} \sim \frac{(m_t)^2}{M_N} \pmatrix{
                      0 &      0      & 0  \cr
                      0 & (m_c/m_t)^2 & 0  \cr
                      0 &      0      & 1  \cr},
\label{M_nu}
\end{eqnarray}
with the MNS neutrino mixing matrix \cite{MNS} defined in the basis where
the charged lepton mass matrix is diagonal,
\begin{eqnarray}
  U_{\rm MNS} &\sim& \pmatrix{
            1/\sqrt{2} & -1/\sqrt{2} & \epsilon    \cr
            1/2        & 1/2         & -1/\sqrt{2} \cr
            1/2        & 1/2         &  1/\sqrt{2} \cr},
\label{U_MNS}\\ \nonumber\\
  \epsilon &=& O(\sqrt{m_c/m_t}).
\end{eqnarray}
This neutrino mass matrix given by Eqs.~(\ref{M_nu}, \ref{U_MNS}) is
nothing but the one used for explaining the atmospheric neutrino 
oscillation\footnote{
From Eq.(\ref{M_nu}), we obtain $m_{\nu_3} \simeq 
\sqrt{\delta m_{\rm atm}^2} \simeq O(0.1)~\EV$.}
and the ``just-so'' oscillation 
of the solar neutrino \cite{BPWW, BGG}.

In this paper we have found a simple SO(10)$_{\rm GUT}$ model which
naturally generates the bi-maximal neutrino mixing suggested from the
atmospheric $\nu_{\mu}$ deficit and the ``just-so'' oscillation solution 
to the solar neutrino problem.
This model maintains the original SO(10)$_{\rm GUT}$ mass relation, 
$m_{\nu_2}/m_{\nu_3} \sim (m_c/m_t)^2$, which is required for the 
``just-so'' scenario \cite{BPWW}.
However, one may think that the present model is already too
complicated, and in this sense the ``just-so'' oscillation seems very
unlikely as stressed by P.~Ramond and one of the authors (T.Y.) 
\cite{MSW_atm_Seesaw}.

Nevertheless, if it turns out to be the case, we will be forced to 
consider drastic changes of underlying physics governs the Yukawa
couplings for quarks and leptons.
We think that our modified SO(10)$_{\rm GUT}$ model presented in this
paper will be a rather mild change among them.

{\bf Note added}

In the text we have restricted our discussion to a specific case of
$f_2 = f_3 =0$ in Eq.~(\ref{mass_matter}).
For a general case ($f_1, f_2, f_3 \neq 0$) we have the following mass
matrix for down-type quarks and charged leptons in the limit $m_u = 0$:
\begin{eqnarray}
  M_{d/l} = m_t \pmatrix{
                    0           & x &                0                  \cr
          (\cos\alpha_1)m_c/m_t & y & (\sin\alpha_1\sin\alpha_2)m_c/m_t \cr
                    0           & z &          \cos\alpha_2             \cr}
          \times\frac{\cos\theta}{\tan\beta},
\end{eqnarray}
which may yield a better fit to the observations.

{\bf Acknowledgments}

T.Y. thanks Hitoshi Murayama for a useful discussion in the early stage
of this work.
Y.N. is supported by the Japan Society for the Promotion of Science.

\newpage

%
%%%%%%%%%%%%%%%%%%%%%%%%%%%%%%%%%%%%%%%%%%%%%%%%%%%%%%%%%%%%%%%
%
% NEW COMMANDS FOR THE BIBLIOGRAPHY
%
%%%%%%%%%%%%%%%%%%%%%%%%%%%%%%%%%%%%%%%%%%%%%%%%%%%%%%%%%%%%%%%
\newcommand{\Journal}[4]{{\sl #1} {\bf #2} {(#3)} {#4}}
\newcommand{\APJ}{Ap. J.}
\newcommand{\CJP}{Can. J. Phys.}
\newcommand{\NC}{Nuovo Cim.}
\newcommand{\NP}{Nucl. Phys.}
\newcommand{\PL}{Phys. Lett.}
\newcommand{\PR}{Phys. Rev.}
\newcommand{\PRep}{Phys. Rep.}
\newcommand{\PRL}{Phys. Rev. Lett.}
\newcommand{\PTP}{Prog. Theor. Phys.}
\newcommand{\SJNP}{Sov. J. Nucl. Phys.}
\newcommand{\ZP}{Z. Phys.}
%%%%%%%%%%%%%%%%%%%%%%%%%%%%%%%%%%%%%%%%%%%%%%%%%%%%%%%%%%%%%%%

%

\begin{thebibliography}{99}
%
\bibitem{Super_K_atm}
        SuperKamiokande Collaboration, 
        Talk by T.~Kajita at {\it Neutrino-98}, 
        Takayama, Japan, June 1998.
%
\bibitem{nu_mu_deficit}
        Kamiokande Collaboration, K.S.~Hirata {\it et al.}, 
        \Journal{\PL}{B280}{1992}{146};\\
        IMB Collaboration, R.~Becker-szendy {\it et al.},
        \Journal{\PR}{D46}{1992}{3720};\\
        SOUDAN-2 Collaboration, W.W.M.~Allison {\it et al.},
        \Journal{\PL}{B391}{1997}{491}.
%
\bibitem{MSW}
        L.~Wolfenstein, 
        \Journal{\PR}{D17}{1978}{2369};\\
        S.P.~Mikheyev and A.~Smirnov, 
        {\sl Yad. Fiz.} {\bf 42} (1985) 1441;
        \Journal{\NC}{9C}{1986}{17}.
%
\bibitem{Hata_Langacker}
        See, for example, N.~Hata and P.~Langacker,
        \Journal{\PR}{D50}{1994}{632}.
%
%
\bibitem{BPW}
        V.~Barger, R.J.N.~Phillips, and K.~Whisnant,
        \Journal{\PR}{D24}{1981}{538}.
%
\bibitem{GK}
        S.L.~Glashow and L.M.~Krauss,
        \Journal{\PL}{B190}{1987}{199}.
%
\bibitem{MSW_atm_Seesaw}
        T.~Yanagida,
        Talk at {\it Neutrino-98}, Takayama, Japan, June 1998;\\
        P.~Ramond,
        Talk at {\it Neutrino-98}, Takayama, Japan, June 1998.
%
\bibitem{Seesaw_models}
        J.A.~Harvey, P.~Ramond, and D.B.~Reiss,
        \Journal{\NP}{B199}{1982}{223};\\
        J.~Sato and T.~Yanagida,
        hep-ph/9710516;\\
        M.~Bando, T.~Kugo, and K.~Yoshioka,
        \Journal{\PRL}{80}{1998}{3004};\\
        C.H.~Albright, K.S.~Babu, and S.M.~Barr,
        hep-ph/9805266;\\
        J.C.~Pati,
        Talk at {\it Neutrino-98}, Takayama, Japan, June 1998;\\
        J.K.~Elwood, N.~Irges, and P.~Ramond,
        hep-ph/9807228.
%
\bibitem{Super_K_sun}
        SuperKamiokande Collaboration, 
        Talk by Y.~Suzuki at {\it Neutrino-98}, 
        Takayama, Japan, June 1998.
%
\bibitem{SNO}
        SNO Collaboration,
        Talk by A.~McDonald at {\it Neutrino-98}, 
        Takayama, Japan, June 1998.
%
\bibitem{democratic}
        M.~Fukugita, M.~Tanimoto, and T.~Yanagida,
        \Journal{\PR}{D57}{1998}{4429};\\
        H.~Fritzsch and Z.Z.~Xing,
        \Journal{\PL}{B413}{1997}{396};\\
        See also, R.~Barbieri, L.J.~Hall, D.~Smith, A.~Strumia, and N.~Weiner,
        hep-ph/9807235.
%
\bibitem{BPWW}
        V.~Barger, S.~Pakvasa, T.J.~Weiler, and K.~Whisnant,
        hep-ph/9806387.
%
\bibitem{Seesaw}
        T.~Yanagida, 
        in {\it Proc. of the Workshop on the Unified Theory and 
        Baryon Number in the Universe}, 
        ed. O.~Sawada and A.~Sugamoto 
        (KEK report 79-18, 1979), p. 95;\\
        M.~Gell-Mann, P.~Ramond, R.~Slansky, 
        in {\it Supergravity}, 
        ed. P.~van Nieuwenhuizen and D.Z.~Freedman 
        (North Holland, Amsterdam, 1979), p. 315.
%
\bibitem{BGG}
        S.M.~Bilenky and C.~Giunti,
        hep-ph/9802201;\\
        A.J.~Baltz, A.S.~Goldhaber, and M.~Goldhaber,
        hep-ph/9806540.
%
\bibitem{Langacker}
        P.~Langacker,
        \Journal{\PRep}{72}{1981}{185}.
%
\bibitem{YY}
        T.~Yanagida and M.~Yoshimura,
        \Journal{\PL}{B97}{1980}{99}.
%
\bibitem{HMY}
        J.~Hisano, H.~Murayama, and T.~Yanagida,
        \Journal{\PR}{D49}{1994}{4966}.
%
\bibitem{Froggatt_Nielsen}
        C.D.~Froggatt and H.B.~Nielsen,
        \Journal{\NP}{B147}{1979}{277}.
%
\bibitem{Future}
        Y.~Nomura, T.~Sugimoto, and T.~Yanagida,
        in preparation.
%
\bibitem{MNS}
        Z.~Maki, M.~Nakagawa, and S.~Sakata,
        \Journal{\PTP}{28}{1962}{870}.
%
\end{thebibliography}
\end{document}